\documentclass[sigplan,10pt]{acmart}

\acmYear{2026}\copyrightyear{2026}
\setcopyright{cc}
\setcctype[4.0]{by}
\acmConference[PACMI '26]{Practical Adoption Challenges of ML for Systems}{September 29--October 2, 2026}{Prague, Czech Republic}
\acmBooktitle{Practical Adoption Challenges of ML for Systems (PACMI '26), September 29--October 2, 2026, Prague, Czech Republic}
\acmDOI{10.1145/3843967.3844678}
\acmISBN{979-8-4007-2998-0/26/09}

\usepackage{microtype}
\usepackage{flushend} 
\usepackage{xcolor}
\usepackage[capitalise,noabbrev]{cleveref}
\usepackage{tikz}
\usetikzlibrary{positioning, calc, arrows.meta}
\usepackage{enumitem}
\usepackage{booktabs}
\usepackage{balance}
\begin{document}

\title{Don't Trust the Code, Check Its Effects}
\subtitle{Runtime Refinement for Regenerated Systems Code Under an Adversarial Generator}

\author{Jinhao Hu}
\affiliation{%
  \institution{Max Planck Institute for Software Systems}
  \city{Saarbr\"ucken}
  \country{Germany}
}

\author{Ashvin Goel}
\affiliation{%
  \institution{University of Toronto}
  \city{Toronto}
  \country{Canada}
}

\author{Laurent Bindschaedler}
\affiliation{%
  \institution{Max Planck Institute for Software Systems}
  \city{Saarbr\"ucken}
  \country{Germany}
}

\begin{abstract}
Recent work uses large language models to generate systems code from
specifications, treating the specification as the durable artifact and the
implementation as disposable. Regenerating the implementation specializes it to
each workload and device. However, that work lives in a
forgiving setting: a
component's externally visible effects, its writes and device commands, are
recoverable, and the generator is honest, so trust is discharged by re-execution.
We target the unforgiving setting: systems code whose effects are
irreversible, produced by a generator that may be adversarial. There, re-execution
cannot check an effect after the fact, and a proof fails silently when its
assumptions do. We take the position that the only safe way to
operate here is to deny the generated code the authority to act. The generated code only plans, while a fixed trusted mediator owns
every effect and performs one only when the specification would have produced it.
Because the guarantee lives in the mediator, not the code, it survives regeneration. We instantiate this as a
reference monitor for regenerated device drivers, and characterize the
\emph{mediability envelope}, six conditions on the effect vocabulary: legibility,
spec-input observability, correlatability, completeness, outcome enumerability, and explicit durability. They decide when
such mediation is possible.
\end{abstract}



\maketitle
\pagestyle{plain}

\section{Generating Systems Components}
\label{sec:motivation}

A growing body of work uses large language models to regenerate systems code on
demand, treating the specification, not the implementation, as the durable
engineering artifact~\cite{Bindschaedler2026ForkOps, Anderson2025SDS,
Liu2026JIT, Stoica2024Specifications, Monperrus2026Bootstrapping,
Mohammadi2026CoherenceDebt}. The pattern
already spans the stack: file systems synthesized from specifications and checked
against specification-derived test suites~\cite{Liu2026SYSSPEC}, distributed
protocols generated with mechanized proofs~\cite{Agarwal2026IDS}, and analytical
database engines synthesized per workload for order-of-magnitude
speedups~\cite{Wehrstein2026BespokeOLAP}. A deployer curates a specification, a
model emits a native implementation, and regenerates it rather than patching it
when the specification changes or a defect surfaces. The specification is what
humans maintain, and the implementation is disposable.

Regeneration buys specialization. Systems components must be tuned to hardware
and workload, and a single implementation commits once to a large design space: a
storage engine's layout, a network stack's congestion control, a driver's
queueing strategy. A model can search that space and re-search it per deployment.
The deployer re-specializes by regenerating, never by editing.

These systems share a setting that makes them tractable: their effects are
recoverable, an analytical query can be re-run, an engine re-benchmarked, a
wrong output recomputed, and their generators are treated as honest. Trust is
therefore discharged by \emph{re-execution}: validate the implementation
against tests or a workload and regenerate on
failure~\cite{Liu2026SYSSPEC, Wehrstein2026BespokeOLAP}.
This works precisely because every effect can be undone.

We target the unforgiving setting, where that assumption fails. Systems code produces
irreversible, externally visible \emph{effects}, the operations through which it
changes state outside itself, such as a device command, a packet send, or a disk
write. Once an effect reaches the host it cannot be recalled, so re-execution
cannot check it after the fact. And the generator cannot be assumed honest: a
compromised provider, a poisoned training set, or prompt injection can emit an
implementation that issues an out-of-spec
effect~\cite{Greshake2023PromptInjection, Chen2026Cordon}. Under irreversible
effects and an adversarial generator, validation by re-execution is unavailable,
and, as we argue next, so is verification alone.

A device driver is the sharpest instance, and our running example. A model
generates the native driver that turns kernel block requests into device commands,
managing queues, DMA, and interrupts, and regenerates it per deployment. Drivers
are a dominant share of kernel code and its faults, tuned across thousands of
devices, and their most dangerous effect, a DMA into arbitrary host memory, is
irreversible and unconfined by construction.

\section{From Verification to Enforcement}
\label{sec:threat}

\Cref{sec:motivation} established the setting; we now pin the threat model. We treat the
generated component as \emph{Byzantine}: it may compute arbitrarily, hide an
out-of-spec effect behind a branch no test exercises, and misreport anything it
tells us. Every value the code supplies is untrusted.

The natural response is to verify the code. Prove that the generated component
refines its specification, admit it, and run it with authority over the device,
as in proof-carrying generation and admission-time
regeneration~\cite{Aggarwal2025AlphaVerus, Yang2025AutoVerus, Liu2026KVerus,
Agarwal2026IDS}. Trust is discharged once, before running, as a property of the artifact.

Verification alone does not survive this setting, because its guarantee is
conditional on assumptions the setting breaks. A proof yields safety only if the
generator produces a checkable proof for every regeneration, the host model it
assumes is complete and sound, the toolchain from proof to binary is trusted, and
the running bits are bound to the verified artifact. An adversarial generator need
not prove anything; a model of real hardware is never complete; compilers and
extraction are rarely verified; and regenerated native code is rarely bound to a
proof. Where any assumption breaks, the proof fails silently and the component
commits an irreversible effect anyway. Verification establishes a property of a
\emph{description}; what reaches the device is a property of the \emph{run}, and
under regeneration the two need not coincide.

Our position is different: trusting the code is not enough, so we deny the
generated code the authority to act in the first place. Safety then no longer
depends on its correctness. The generated component becomes a \emph{planner} that
proposes effects but performs none; a fixed, handwritten \emph{mediator} owns every
\emph{effect primitive}, the low-level operations through which state leaves the
machine, and performs one only after checking, against the specification and its
own state, that the effect is one the specification would produce. The guarantee
no longer rests on the code but on the mediator, which observes the actual effect
at the one point it can still be refused, from a bounded abstract state, not the
device's internals.

\looseness-1 This trades a verification toolchain for a mediator that carries its own
abstract state, its own account of what is safe, and its own compiler. We do
not claim that the trusted base is smaller, and size is not the useful comparison.
What separates them is the work a regeneration forces. A proof binds one
binary, so every regenerated implementation must earn its evidence again, from
proof through extraction to binary. Our compiler checks the specification,
which does not change when the implementation does, so a regenerated driver
re-runs nothing and gains no authority.

\looseness-1 This is not a claim that verification cannot be safe; a stack
verified down to a trusted machine model needs no mediator. Under an
adversarial generator, native code, and irreversible effects, though,
verification's preconditions are unmet, so the trusted decision must be made
at the effect, not the artifact. Where achievable, verification raises
availability by making proposals usually correct, but safety rests on
enforcement.

We make three contributions. We recast trust for regenerated systems code as
\emph{effect authority} rather than code verification: the code only plans, and a
fixed mediator owns and adjudicates every effect. We instantiate this as a
reference monitor for regenerated drivers~(\cref{sec:boundary}), extending
driver-safety monitoring~\cite{Williams2008RVM} to functional refinement. And we
characterize the \emph{mediability envelope}~(\cref{sec:envelope}): six conditions
on the effect vocabulary that decide when such mediation is possible. This is a
position paper: no formal semantics, mechanized proof, implementation, or
measurement.

\section{The Boundary, by Example}
\label{sec:boundary}

\Cref{fig:architecture} illustrates the boundary we call \emph{proposal-commit
separation}: the driver \emph{proposes} device commands, and the mediator commits
one only when it refines the specification.

\begin{figure}[h]
\centering
\begin{tikzpicture}[
  font=\footnotesize\sffamily,
  box/.style={draw, line width=0.8pt, rounded corners=3pt, minimum height=7.5mm,
              align=center, inner sep=1mm},
  arrow/.style={-{Stealth[length=2mm]}, line width=0.8pt, draw=black!80},
  plabel/.style={font=\scriptsize, fill=white, inner sep=1pt, align=center},
]
\definecolor{untrustedline}{HTML}{D55E00}
\definecolor{untrustedfill}{HTML}{FBEADE}
\definecolor{trustedline}{HTML}{0072B2}
\definecolor{trustedfill}{HTML}{E1EEF6}
\node[box, draw=untrustedline, fill=untrustedfill, minimum width=26mm] (handler)
  {\textbf{generated driver}\\\scriptsize untrusted, confined\\\scriptsize plans; cannot act};
\node[box, draw=trustedline, fill=trustedfill, right=13mm of handler,
      minimum width=29mm] (mediator)
  {\textbf{mediator} {\scriptsize(trusted)}\\\scriptsize spec + abstract state\\\scriptsize sole effect authority};
\node[box, draw=black!55, fill=black!6, minimum width=30mm]
  at ($(mediator.north)+(0,8mm)$) (kernel) {kernel block layer};
\node[box, draw=black!55, fill=black!8, below=6.5mm of mediator, minimum width=29mm]
  (device) {device (MMIO, DMA)};
\draw[arrow] ($(kernel.south)+(-6mm,0)$) --
  node[plabel, pos=0.5, left=1mm]{request} ($(mediator.north)+(-6mm,0)$);
\draw[arrow] ($(mediator.north)+(6mm,0)$) --
  node[plabel, pos=0.5, right=1mm]{validated\\result} ($(kernel.south)+(6mm,0)$);
\draw[arrow] ([yshift=4.5mm]mediator.west) --
  node[plabel, above=0.3mm]{\texttt{Dispatch}} ([yshift=4.5mm]handler.east);
\draw[arrow] ([yshift=0.5mm]handler.east) --
  node[plabel, above=0.3mm]{\texttt{Observe}, \texttt{Propose}} ([yshift=0.5mm]mediator.west);
  \draw[arrow] ([yshift=-3.5mm]mediator.west) --
    node[plabel, pos=0.5, below=0.3mm]
      {\texttt{Commit} $\mid$ \texttt{Reject}}
    ([yshift=-3.5mm]handler.east);
\draw[arrow] ([xshift=-4.5mm]mediator.south) --
  node[plabel, pos=0.62, left=1mm]{effect\\primitives} ([xshift=-4.5mm]device.north);
\draw[arrow, dashed] ([xshift=4.5mm]device.north) --
  node[plabel, right=1mm]{completion\\{\itshape(env.)}} ([xshift=4.5mm]mediator.south);
\draw[dashed, line width=0.6pt, black!65]
  ($(handler.north east)!0.5!(mediator.north west)+(0,3mm)$) --
  ($(handler.south east)!0.5!(mediator.south west)+(0,-4mm)$);
\end{tikzpicture}
\caption{Proposal-commit boundary for a driver. The kernel's request
enters the mediator, which dispatches to the confined driver; the driver may only
\texttt{Observe} and \texttt{Propose}. The mediator alone holds the specification
and effect primitives (MMIO, DMA), committing only effects that refine it.}
\Description{A kernel request enters a trusted mediator, which dispatches it to
a confined generated driver. The driver exchanges Observe/Propose and
Commit/Reject messages with the mediator. The mediator alone connects to
the device via effect primitives, receives completions, and returns validated
results to the kernel.}
\label{fig:architecture}
\end{figure}
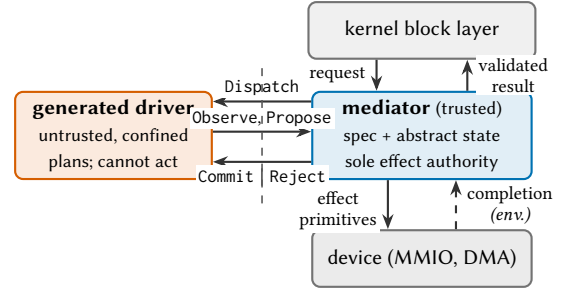

\begin{figure}[t]
\begin{center}\begin{minipage}{0.88\columnwidth}\hrule\vspace{2pt}
\begin{footnotesize}\begin{verbatim}
operation WRITE(lba, len, buf)  -- kernel's request
  observe  queue.free, dma.grant[buf]
  permit   op   = WRITE        -- fixed by request
           lba  = req.lba
           len  = req.len
           dma  = req.buf
           slot in queue.free  -- driver's choice
  outcome  ok       -> req.done
           EIO      -> req.failed, device unchanged
           short(n) -> req.partial(n)
\end{verbatim}\end{footnotesize}
\vspace{1pt}\hrule\end{minipage}\end{center}
\caption{The \texttt{WRITE} clause of a driver specification. The mediator's
abstract state is the outstanding requests, the queue's free slots, and the DMA
grants. \texttt{permit} separates the fields the specification fixes from the
pending request, which the mediator re-derives, from the one field
the driver is free to choose. \texttt{outcome} enumerates what the primitive
may return and the abstract-state transition each produces.}
\Description{A specification fragment for a block-device WRITE operation,
listing the abstract state it observes, which command fields are fixed by the
request and which the driver may choose, and the three outcomes the primitive
may return with their state transitions.}
\label{fig:spec}
\end{figure}

\paragraph{The specification.} The mediator holds a specification of the
driver's \emph{interface}, not of its implementation. \Cref{fig:spec} gives the
\texttt{WRITE} clause: a small, device-generic transition relation over the
outstanding requests and the queue state. It is a fraction of the driver's
queueing and DMA machinery, so checking a command against it costs far less
than producing one. Runtime values are taken from the request itself.
\texttt{req.lba} names whatever the kernel asked for in the request now
pending, and the mediator, which holds that request, re-derives the value
rather than taking the driver's word for it. The driver chooses only the queue
slot, and the specification bounds that choice to the slots the driver observes free.

\paragraph{The protocol.} The driver never touches the device. The mediator
\texttt{Dispatch}es a pending request. While planning, the driver must first
\texttt{Observe} the abstract state the specification
permits it to see, the free queue slots and DMA grants of \cref{fig:spec}; it may observe again before proposing.
\texttt{Observe} returns state, never authority, together with a fresh \emph{sealed
token}, a value the driver cannot forge, binding the request, the versions of the
abstract state that were read, and an expiry. The driver necessarily
plans against a snapshot, and that snapshot may move before it proposes; the mediator rechecks the token during \texttt{Validate}, under the commit
lock. A later observation invalidates the preceding token, so a proposal always describes a plan against one identified observation.

\texttt{Propose} is the driver's only message that asks the mediator to
perform an effect. It submits an \emph{attempted effect}, a device command, and
nothing more, together with its token. Every field \emph{of the proposal} is an untrusted
witness. The driver says nothing about what the effect will return, because the
specification has already said it: \cref{fig:spec} closes the outcome set at
three entries for \texttt{WRITE} and pairs each one with its abstract-state
transition. Once the primitive has run, the mediator therefore needs nothing
further from the driver.

\looseness-1 The mediator runs \texttt{Validate} under a
commit lock, and rejects a replayed, expired, superseded, or stale token. It
then checks that every field the
specification fixes matches what it re-derives from the pending
request, which it holds itself and never reads from the proposal, and that each freely chosen field falls inside the permitted set.
\texttt{Validate} consumes the token whether it accepts or rejects the
proposal: one request cannot produce a second effect by replaying a token. 

Only then does the mediator
\texttt{Commit}. It invokes the trusted primitive and reads the \emph{actual}
outcome, which \cref{fig:architecture} shows returning as \emph{completion}. The
specification, not the driver, says which transition that outcome produces, and
the mediator records it and delivers the validated result to the kernel.
A device error that the specification lists, such as the \texttt{EIO} of
\cref{fig:spec}, is a committed outcome rather than a failure of mediation. The
mediator records \texttt{req.failed} and reports the error upward. When
\texttt{Validate} fails instead, it \texttt{Reject}s before any irreversible
primitive runs, so a rejected proposal is externally silent.

\paragraph{Catching a wrong write.} \looseness-1 The
kernel issues \texttt{write LBA 100..107 from $B$}; consider a driver that proposes
\texttt{\{op=WRITE, LBA=200, len=8, src=$B$, slot=$k$\}} instead of the
requested LBA 100:
it overwrites the wrong block with $B$'s contents, an effect no later regeneration can undo.
Every other check passes: the DMA source is still the authorized buffer $B$,
so an IOMMU, which never sees an LBA at all, has nothing to object to;
200 is a legal LBA, so a device-safety monitor that never observes the kernel's request, such as a reference validation mechanism~\cite{Williams2008RVM},
has no requested value to compare against; an LBA the test suite never
exercised would pass sampled admission validation~\cite{Liu2026SYSSPEC} just the same.
Only a check that holds the pending request and compares it against the proposed command catches this:
the mediator's \texttt{Validate} rejects it before the doorbell rings,
because $100 \neq 200$, and nothing is written to the device.
This same comparison covers the rest of the proposal. A wrong \texttt{len}, or a
DMA source other than $B$, fails identically, since \cref{fig:spec} fixes both
to the request. A \texttt{slot} outside the observed free set fails instead
against the permitted set, and a proposal naming no pending request is
rejected at the token. Every \texttt{Propose} carries the token from its most
recent \texttt{Observe}, and \texttt{Validate} consumes it, so a request in
flight yields at most one effect, however often the driver replans, and a
driver that proposes nothing stays silent. Absent a crash, one request never
yields a second effect.

\paragraph{Confinement and ownership.} The driver runs with no direct kernel
access, no device register mapping, and no DMA authority; the mediator is its
only outward channel. Confinement blocks any effect the driver might attempt on
its own, rather than relying on inspect-and-continue \texttt{seccomp}
notification, which is unsafe against concurrently mutated
arguments~\cite{Kerrisk2026SeccompNotify}. The mediator owns the trusted effect
primitives, observation-token issuance, DMA authorization through the IOMMU,
the abstract state of \cref{fig:spec} and the commits that advance it, the
append-only log in which each committed effect is recorded together with the
outcome it produced, and the specification interpreter. It is the authority for
the declared effect vocabulary.

\paragraph{Concurrency.} Drivers may plan and compute concurrently, but the
mediator serializes the parts that matter: validation, primitive invocation,
abstract-state update, and log insertion run under a single commit lock, tokens
are revalidated after the lock is taken, and a stale proposal is rejected and
replanned. The commit-log order is the specification's serialization order.
Serializing primitives does not make a multi-primitive guest operation atomic;
an operation that needs cross-effect atomicity requires one atomic primitive,
substrate locking, or exclusion from the vocabulary.

\paragraph{Guarantee.} A realization should claim
\emph{committed-trace safety}: every committed sequence of externally visible
effects is a prefix of a trace the specification allows over the declared effect
vocabulary. It presupposes that the effect vocabulary lies within the
mediability envelope~(\cref{sec:envelope}), and it is conditional on a correct
specification and interpreter,
complete mediation with no bypass, trusted effect primitives, serialized commits,
outcome enumerability, and a stated crash model. The guarantee excludes liveness
and availability, since a Byzantine driver may propose nothing or select the
worst permitted proposal, and it does not cover timing channels or defects in
the specification itself.

\section{The Mediability Envelope}
\label{sec:envelope}

A specification is written against an interface: it names that interface's
operations, such as \texttt{mkdir(path,mode)} or \texttt{write LBA 100..107}, and
the objects those operations touch, such as paths, credentials, queue slots, and
DMA buffers. We call that interface's level of abstraction the specification's \emph{altitude}.
The architecture works only where the effect vocabulary lets the mediator see and
check the right thing cheaply, at that altitude, without becoming the code it checks. Refinement
is a relation over what the mediator must observe and hold against the
specification across the life of a request: the \emph{intent} (the requested operation), 
the \emph{delta} (what actually changed), the \emph{result} (what the caller is told),
and, over time, whether what was recorded survives a crash. Whether a
vocabulary supplies these, cheaply and independently, defines a \emph{mediability envelope}. 
Six conditions delimit it, and they are of two kinds. The first three ask what
the effect vocabulary supplies, and altitude settles them, so they decide
whether a domain can be mediated at all. The last three ask what the deployment
supplies, and engineering settles them at any altitude. We read each condition against
two vocabularies: the driver's device commands, which satisfy it, and a file
system's block writes, which serve as a foil for what failure looks like. Where a condition fails, the mediator can regain it
only by reconstructing what the vocabulary discarded.

\paragraph{What the vocabulary must supply.}
\looseness-1 \emph{Legibility.} A single effect must be legible, on its own, as an instance
of one operation: from the effect alone, without replaying what came before,
the mediator must be able to tell which operation it belongs to.
A device command names its operation: \cref{fig:spec}'s \texttt{WRITE} is legible
on its own, distinguishable from a \texttt{READ} by its own encoding. A block write does not: the
same write to a directory block and a bitmap could be a \texttt{mkdir}, an
\texttt{rmdir}, a rename, or a corruption, so whether an operation changed
state, changed nothing, or did the opposite is invisible.
Ambiguity denies the mediator the intent.

\emph{Spec-input observability.} The mediator must observe every input the
transition reads. \Cref{fig:spec}'s \texttt{WRITE} observes
\texttt{dma.grant[buf]}, the caller's own authorized source buffer, before
permitting a \texttt{slot}; \texttt{mkdir} similarly is licensed only against
the caller's credentials, but the virtual-filesystem interface carries
them while the block layer strips them, so a permission-dependent transition
cannot be checked there.

\looseness-1 \emph{Correlatability.} Effects must group into per-operation
transactions with a recoverable order under concurrency. At the specification's
altitude a device command is self-delimiting, so grouping is free: one proposal,
one transaction, serialized at commit. \Cref{fig:spec}'s \texttt{WRITE} is
exactly this. Block writes are not: concurrent
operations interleave unlabeled writes over shared structures, so they cannot be
grouped by address, and no serialization recovers an identity the vocabulary
never carried. Even where effects self-delimit, physical write order can diverge
from commit order, and only the commit lock recovers it. Certifying a stream that
fails either half means modeling the component's own concurrency control.

\paragraph{What the deployment must supply.}
\emph{Completeness.} With intent legible, the record of what changed must be kept
sound: every externally visible state change must cross
the boundary, as a validated effect primitive or a mediator-owned environment
transition; any third path voids the guarantee. The standing risk is effect
authority the component already holds: a writable shared mapping or DMA lets it
change host-visible state before any proposal. Both must be excluded or reified
into proposals, shared memory through shadow pages, DMA through IOMMU-authorized
buffers. \cref{fig:spec}'s \texttt{dma.grant[buf]} is exactly this
reification, checked before any \texttt{WRITE} is permitted, not assumed.

\emph{Outcome enumerability.} The record must also be trustworthy on the way out:
the outcomes a primitive may produce, as reported at the mediator's interface,
must be closed and enumerable, so that whatever it returns, success, a device
error, or a partial completion, can be classified against the specification's
allowed set the moment it returns, before any abstract state is advanced, as
\cref{fig:spec}'s three-entry \texttt{outcome} clause does for \texttt{WRITE}.
The mediator needs no model of why the device produced an
outcome, only an enumeration of the forms it can take; an unenumerable
outcome space forces the mediator to guess at commit time what an incomplete
device model forces a prover to guess at admission time.

\emph{Explicit durability.} Persistence extends the same concern into time:
whether what the record says happened still holds after a crash.
\Cref{fig:spec}'s \texttt{WRITE} illustrates the gap: its \texttt{ok} outcome
means the bytes reached the device, not that they survive a crash. Durability must
be a first-class effect the component proposes, a flush or barrier, not a property
that emerges from write ordering, since a crash truncates the stream at an
arbitrary point while destroying volatile state. If it is explicit, the mediator
checks that a claimed-durable \texttt{WRITE} issued a following flush the device
confirmed; if implicit, the mediator must reconstruct the component's barrier
discipline and keep its own crash-consistent store.

\looseness-1 One condition straddles the split. Correlatability's grouping half
belongs to the vocabulary, but its ordering half is closed only by the commit
lock. So the rule narrows: mediate at the specification's own
interface, never below it, and treat confinement, serialization, outcome
closure, and durability as separate engineering. Below that altitude the mediator must
\emph{lift} effects back to operations, re-implementing the read path, concurrency
control, and recovery, and re-attaching the intent and authority the vocabulary
discarded.

Within the envelope, a separate question is economic: how much standing state a
check must consult. A driver sits at the cheap end (queue descriptors, DMA
grants); a file system at the virtual-filesystem interface sits at the costly end,
owning the namespace and per-file digests while the component owns the physical
data plane; arbitrary specialization or bugs then degrade availability, not
silently corrupt data. At the block layer the same system falls out of the
envelope: legibility, spec-input observability, and correlatability fail. Altitude is
the line between checking and re-implementing.

\section{Related Work}
\label{sec:related}

\looseness-1 The approaches we reject have deep lineages.
\Cref{tab:design-space} organizes the design space along two questions: when
the check runs, at admission or at commit, and what it checks, permission or
correctness. \emph{Proof} certifies the artifact
at admission, before it runs, from proof-carrying code~\cite{Necula1997PCC} to
verified systems~\cite{Klein2009SeL4, Lattuada2023Verus} and provers aimed at
generated code~\cite{Yang2025AutoVerus, Aggarwal2025AlphaVerus, Liu2026KVerus,
Agarwal2026IDS}. The proof is re-earned for every regeneration. \emph{Policy}
withholds effect authority and asks only whether an effect is permitted: the
eBPF verifier~\cite{Gershuni2019EBPF} checks at admission, while software fault
isolation~\cite{Wahbe1993SFI} and NaCl~\cite{Yee2009NativeClient} check at
commit. Policy substrates such as~\cite{Young2019GVisor, Agache2020Firecracker, WASISubgroup2026WASI}
can also interpose at that point, but typically decide whether a primitive,
fd, address, or DMA buffer is authorized, not whether it refines the current
trusted request and contract state; a policy that retains and checks the
specification's transition relation there is already a mediator in this
paper's sense.

\looseness-1 Our mediator composes established mechanisms: confined code that
delegates to a trusted agent~\cite{Garfinkel2004Ostia, Kerrisk2026SeccompNotify},
opaque handles from capability systems~\cite{Hardy1985KeyKOS, Shapiro1999EROS,
Watson2010Capsicum}, speculate-and-commit from system
transactions~\cite{Porter2009TxOS, Jana2011TxBox}, and behavioral-model checking
from model-carrying code~\cite{Sekar2003MCC}, edit
automata~\cite{Schneider2000Enforceable, Ligatti2005EditAutomata}, and shield
synthesis~\cite{Bloem2015Shield}. Runtime verification compiles specifications
into monitors that run alongside the
program~\cite{Leucker2009RuntimeVerification, Chen2007MOP}, and runtime
refinement checking is two decades old~\cite{Elmas2005VYRD}, but both observe a
trace rather than gating it. Output-buffering speculation defers an external
effect until it is known safe~\cite{Nightingale2005Speculator,
Nightingale2006RethinkSync}, but it presumes a recoverable substitute stands in
for the effect; it cannot withdraw a command the device has already run.

The closest work is driver safety through a reference validation
mechanism~\cite{Williams2008RVM}: untrusted, restartable native drivers, each
interaction checked at commit by a specification-compiled monitor that survives
replacement. That monitor enforces a device-safety automaton; ours checks
functional refinement against a generator that rewrites the driver on demand.
Runtime checking of file-system metadata at commit is a similar line: Recon and
its successor check consistency and atomicity invariants over block
writes~\cite{Fryer2012Recon, Fryer2014Integrity}, but below the specification's
altitude, reading only metadata and never the caller's request, so they catch
structural corruption, not permitted-but-wrong outcomes. Concurrent work mediates
LLM \emph{agents} at the tool boundary~\cite{Chen2026Cordon, Moon2026Containment,
He2026SAB, SantosGrueiro2026Portico, Fernandez2026ACP, He2026PDD,
Mohammadi2026Atomix}, but assumes an
honest generator and acts above regenerated native systems code. Our companion
workshop papers place a mediation boundary around agent
explorations~\cite{Hu2026ExternalizationBarriers} and price fleets' residual
risk~\cite{Mohammadi2026IrreversibilityBudget}, but target agent workloads, not
regenerated systems code.

\begin{table}[h]
\scriptsize
\setlength{\tabcolsep}{3pt}
\renewcommand{\arraystretch}{1.05}
\begin{tabular}{@{}p{0.23\columnwidth}p{0.3\columnwidth}p{0.33\columnwidth}@{}}
\toprule
& \textbf{Permission}\newline(effect is allowed)
& \textbf{Correctness}\newline(effect refines the spec) \\
\midrule
\textbf{At admission}\newline per artifact; redone on regeneration
  & static allowlists; the eBPF verifier
  & proof-bearing generation (PCC, Verus, IDS); admission validation (SYSSPEC) \\
\textbf{At commit}\newline per effect; survives regeneration
  & policy substrates (NaCl, gVisor, WASI, seccomp); device-safety monitors (RVM)
  & \textbf{runtime refinement (this paper)} \\
\bottomrule
\end{tabular}
\caption{Design space for trusting regenerated systems code. Admission-time
checks are redone on regeneration; commit-time checks survive in the mediator.
This paper claims the fourth cell.}
\label{tab:design-space}
\end{table}

\section{What a Realization Must Answer}
\label{sec:agenda}

\looseness-1 A realization must answer four engineering questions beyond mediability.
\emph{Cost}: how large is the mediator's trusted code and abstract state against
the component it guards, what latency does serialized commit add, and how often do
proposals go stale? If the mediator grows with the
implementation, the envelope was drawn at the wrong altitude. Whether drivers clear that bar is a fair doubt: the per-effect
check is cheap only because the driver's abstract state stays small, and thin
margins argue for regenerating device by device. \emph{Adapter
trust}: \looseness-1 each effect primitive reaches the device through an \emph{adapter}, the
trusted code that turns a validated command into the register
writes and DMA descriptors the hardware expects. Is each adapter mechanically
verified, audited against a contract, or excluded from the trusted base?
\emph{Availability}: a Byzantine component can propose nothing, stall with
fresh observations, or exhaust its permitted choices, so a deployment needs a
deadline, budgets, or a fallback.
\emph{Unknown unknowns}: a specification defect is enforced faithfully, and an
unenumerated outcome breaks outcome enumerability, so a realization must fail
stop on both.

\looseness-1 Regenerated systems code needs a trust boundary that survives
regeneration. Denying it effect authority and mediating each effect provides
one, and the envelope says when: this paper isolates the laws a regenerable OS
builds on.

\begin{acks}
We thank the PACMI '26 reviewers for their comments. We used AI tools for
editorial assistance (drafting, copy-editing, and trimming for length); all
technical content, results, and conclusions are our own.
\end{acks}

\balance
\bibliographystyle{ACM-Reference-Format}
\bibliography{main}

\end{document}